\documentclass{article}
\usepackage{amsmath}
\usepackage{amsfonts}
\usepackage{amssymb}
\usepackage{graphicx}
\newtheorem{theorem}{Theorem}

\newtheorem{proposition}[theorem]{Proposition}

\begin{document}

\title{The formation of share market prices \\under \\heterogeneous beliefs and common knowledge}

\author{Yuri Biondi$^1$\thanks{ yuri.biondi@polytechnique.edu}, Pierpaolo Giannoccolo$^2$\thanks{pierpaolo.giannoccolo@unibo.it}, Serge Galam$^3$\thanks{serge.galam@polytechnique.edu}\\ \\$^1$P\^ole de Recherches en \'Economie et Gestion (PREG),\\
\'Ecole Polytechnique and CNRS, \\Boulevard Victor, 32,
F-75015 Paris, France\\$^2$Department of Economics. University of Bologna, 40100 Bologna, Italy,\\
$^3$Centre de Recherche en \'Epist\'emologie Appliqu\'ee (CREA),\\  \'Ecole Polytechnique and CNRS, \\Boulevard Victor, 32,
F-75015 Paris, France}
\date{May 17, 2011}
\maketitle

\begin{abstract}
Financial economic models often assume that investors know (or agree on) the
fundamental value of the shares of the firm, easing the passage from the
individual to the collective dimension of the financial system generated by
the Share Exchange over time. Our model relaxes that heroic assumption of one
unique "true value" and deals with the formation of share market prices
through the dynamic formation of individual and social opinions (or beliefs)
based upon a fundamental signal of economic performance and position of the
firm, the forecast revision by heterogeneous individual investors, and their
social mood or sentiment about the ongoing state of the market pricing
process. Market clearing price formation is then featured by individual and
group dynamics that make its collective dimension irreducible to its
individual level. This dynamic holistic approach can be applied to better
understand the market exuberance generated by the Share Exchange over
time.

\end{abstract}



\section{Introduction}

Advances in heterogeneous agents modeling from economics \cite{hom, cris}  and complex systems dynamics in sociophysics \cite{chakra, sama} call for an understanding of the working of the financial market based upon the collective and dynamic properties of systems featured by interacting parts and structures. These elements can be atoms or macromolecules in a physical context, as well as people, firms or regulated Exchanges in a socio-economic context. These approaches aim then to analyze the properties of socio-economic systems over time by focusing on interactions, relationships and the overall architecture of them.  

Drawing upon these advances, this paper integrates the phenomenon of opinion dynamics studied by sociophysics \cite{fortunato-review, galam-review} to an economic dynamic model of market price formation over time through hazard and interaction \cite{bulle}. The study of opinion dynamics has been a long and intensive subject of research among physicists working in sociophysics \cite{martins, ausloos-religion,  pierluigi,  schneider, sznajd, deffuant, pair}: we apply here the Galam sequential probabilistic majority model of opinion dynamics \cite{pair, mino, hetero}.

During the last decades, financial market analysis has assisted to the
proliferation of financial economic models that relax received assumptions of
full knowledge, individual rationality and market efficiency. However, many
models remain somewhat tied to an equilibrium approach to the formation of
share market prices over time. This approach entails a pricing rule that
satisfies all the market orders simultaneously passed by all investors in the
purpose to maximize their expected utilities. This approach actually implies a
peculiar understanding of the market coordination between individual
investors. This coordination is supposed to be achieved in a solitary moment
beyond time and context \cite{Shubik} when all investors contemplate the past,
present and future of the business firm and univocally agree on its
fundamental value of reference. 

Once this unanimous consensus achieved, they
perform market transactions at that price, which does not change unless the
fundamental value of the firm does change \cite{Tirole}. Therefore, the share
market price is supposed to incorporate (all the available information on) the
fundamental value of the firm at every instant \cite{Fama2,  Fama3}. The
share market price becomes a sufficient statistics of the fundamental value of
the firm \cite{Fama1}, and investors are then supposed to know (or agree with)
the fundamental value of its shares, even though the current market price may
diverge from this "true value" in some ways over time. The understanding and
the modeling of market pricing, and the dynamics of individual and collective
opinions, are then driven by this assumption of uniqueness of the value of the firm.

Our model relaxes this heroic assumption of the market price as the best
evidence of the "true value," and deals with the formation of share market
price of one firm through the dynamic formation of individual and social
opinions (or beliefs) based upon a fundamental signal $F_{t}$ on the economic
performance and position of the firm, the market clearing price of each share
$p_{t}$, and a social mood (or sentiment) $m_{t}^{{}}$ on the ongoing state of
market pricing process. Accordingly, individual investors are assumed to form
their personal opinions - which orient their financial decisions of sell or
hold, and buy or wait - in a fundamentally interactive context  \cite{Phelps}.
At every instant $k$, each investor $i$ does form its opinions respectively on
the evolution of corporate fundamentals and the market clearing price that is
continuously changed by achieved transactions through the Share Exchange.

Nothing can assure one investor about the permanent alignment between his
opinion on the evolving fundamentals, its opinion on the current market price,
and the market price itself \cite{Frydman}; nor can he be sure that the market
order - which he passes through the Share Exchange according to those opinions
- may be eventually satisfied.\ In this dynamic setting, the formation of
share prices critically depends on both the interactive formation of social
opinions among investors, and their common knowledge of corporate fundamentals
over time. Every investor strives then to revise its price expectations
$\left.  E_{t}(p_{t+1})\right\vert _{i}$ according to the dynamics of the
fundamental signal $F_{t}$ and the social market sentiment $m_{t}^{{}}$.

\section{Definition of Variables and Timing}

The formation of share market price over time depends here on the dynamic
formation of individual and social opinions (or beliefs) based upon a
fundamental signal $F_{t}$ on the economic performance and position of the
firm, the market clearing price of each share $p_{t}$, and a social mood (or
sentiment) $m_{t}^{j}$ about the market pricing. These three dimensions (or
layers, or orders) correspond to three different rhythms of change, that is,
three different timings:

\begin{itemize}
\item $F_{t,h}$, the fundamental signal, has the slowest rhythm or the largest
lag (duration). This means that $F_{t}$ can be constant for $t+h$ periods; it
lasts for $h$ periods;

\item $p_{t}$, the market clearing price, when exist, changes at each period
$t$;

\item $m_{t,k}^{j}$, the social mood, has the quickest rhythm or the shortest
lag. At each period $t$, its value is the final result of $k$ interactions;
each mood lasts indeed for $\frac{1}{k}$ periods.
\end{itemize}

Two distinctive forces drive then the market clearing price formation trough
time.\ From one side, ongoing market pricing is submitted to individual
guesses and intentions, hopes and fears, subsumed by the social mood
$m_{t,k}^{j}$ and its quickest interactions; from another side, it is
concerned with the slowest history of reporting and disclosure that, in
principle, may be partly public, consistent, and conventionally agreed. This
general \textit{system} (which is no longer an equilibrium)\footnote{Our
analysis distinguishes system and equilibrium as distinctive concepts.}
consists in and depends upon the coherence and universal diffusion of relevant
and reliable knowledge through a price system (providing market information)
and an accounting system (disclosing firm-specific, fundamental information)
publicly determined and announced.

In particular, the \textbf{fundamental} \textbf{signal }is assumed to be
common knowledge among all investors:

\begin{itemize}
\item $F_{t}$ is the fundamental signal about the economic performance and
position generated by the business firm over time; it is fundamentally related
to the firm's share price, but agents do not know (or agree on) the working of
this relationship;

\item $F_{t}$ can be positive or negative and is exogenous to the model;

\item Each agent applies an individual weight $\varphi_{i}\in\left[
0;1\right]  $ to this signal, related to its personal confidence degree on it,
from $\varphi_{i}=0$ (no confidence at all) to $\varphi_{i}=1$ (full
confidence); this implies that all agents agree on the direction (sign) of the
fundamental signal, but disagree on its material impact on the share price.

\item In some specifications of the model, $F_{t}$ may influence the social
mood $m_{t,k}^{j}$.
\end{itemize}

The \textbf{social} \textbf{mood} (or market sentiment) captures the group
interaction that generates the collective opinion on the current state of
market pricing\textbf{:}

\begin{itemize}
\item $m_{t,k}^{j}\in\left[  0;1\right]  $ is the mood of group $j$ at time
$t$, resulting from $k$ group interactions (steps) starting from
$m_{t,k=0}^{j}$;

\item At each time $t$, $m_{t,k=0}^{j}\in\left[  0;1\right]  $ exists and is
exogenous to the model;\ in fact, $m_{t,k=0}^{j}$ may be endogenous to the
model; in particular, it may depends on $F$.
\end{itemize}

The \textbf{market clearing price} (when exists) is generated by the matching
of aggregate supply and demand, which are based upon heterogeneous price
expectations by individual agents\textbf{:}

\begin{itemize}
\item $p_{t}$ is the market clearing price at time $t$;

\item By assumption, $p_{t}\geq0$;

\item $\left.  E_{t}(p_{t+1})\right\vert _{i}^{j}$ is the price expectation at
time $t$ by agent $i$ belonging to the group $j$ on the market clearing price
at period $t+1$.
\end{itemize}

\textbf{Individual} \textbf{investors} have both group and individual
heterogeneities regarding the formation of their expectations, which are then
based upon individual and social opinions (or beliefs).\ In particular:

\begin{itemize}
\item Investors are distinguished between actual and potential
shareholders.\ Analytically, they belong then to two groups $j=S,D$, where $S$
denotes supply by potential sellers (actual shareholders), while $D$ denotes
demand by potential buyers (potential shareholders);

\item In each group $j$, the number of agents is normalized to one, with
$i\in\left[  0;1\right]  $;

\item In each group $j$, every agent $i$ is characterized by an individual
weight $\varphi_{i}\in\left[  0;1\right]  $ that is applied to the fundamental
signal $F_{t}$;

\item In each group $j$, agents are further characterized by the social mood
$m_{t,k}^{j}$ that constitutes the market sentiment expressed by group $j$ at
time $t$; its weight results from $k$ inter-individual interactions between
$t-1$ and $t$.
\end{itemize}

\section{The formation of individual expectations}

Following \cite{Hirota} and \cite{Heemeijer}, every agent forms its price expectation according to the
following generic function:

\begin{equation}
\left.  E_{t}(p_{t+1})\right\vert _{i}^{j}=p_{t}+m_{t,k}^{j}\left(
p_{t}-p_{t-1}\right)  -\beta_{i}^{j}\left(  \left.  E_{t-1}(p_{t})\right\vert
_{i}^{j}-p_{t}\right)  +\gamma^{j}\varphi_{i}F_{t}
\end{equation}

with $j=S$ (Supply), $D$ (Demand); $i,\varphi_{i}\in\left[  0,1\right]  $;
$m_{t,k}^{j}\in\left[  0,1\right]  $; $\beta_{i}^{j}\in\left[  0;1\right]  $;
$\gamma^{j}>0$, and

\[
\varepsilon_{i,t}^{j}\equiv\left(  \left.  E_{t-1}(p_{t})\right\vert _{i}
^{j}-p_{t}\right)  \text{.}
\]

This equation comprises forth elements.\ The first element is the past
clearing price $p_{t}$.\ The second element is the market signal (or price
trend) that is weighted by the social opinion $m_{t,k}^{j}$ of the group $j$
at time $t$, expressing the group's ongoing market confidence. The third
element is the individual forecasting revision that consists of the difference
between the past price expectation and the current realized price. This
revision is weighted by $\beta_{i}^{j}$\ which may include both group and
individual heterogeneities.\ The forth element denotes the formation of an
individual opinion by investor $i$ (belonging to the group $j$) based upon
available fundamental information $F_{t}$, which is common knowledge for both
groups and all the individual investors, and is weighted then by the
individual parameter $\varphi_{i}$. This structure of individual expectations
follows the dual structure which the share market process is embedded in: From
the cognitive viewpoint, investors are confronted with fundamental information
from the business firm (they invest in) from one side, and the market pricing
from another side.\ From the financial viewpoint, they are confronted with
dividends and earnings generated by the business firm, and the capital gains
and losses involved in the market trading (see \cite{Biondi} for
further details). The firm side is subsumed here by the factor $F$, while the
market side is captured by the price trend $_{t-1}\Delta_{t}\left(  p\right)
$.

Following the Galam's specification of the formation of social opinions \cite{mino, hetero}, we
can define the generic function of the social mood $m_{t,k}^{j}$ as follows:

\[
m_{t,k}^{j}=f\left(  m_{t,k=0}^{j},k_{t}^{j},F_{t},F_{t-1}\right)
\]

where the fundamental signal $F$ can influence $k_{t}^{j}$. For each group
$j=S,D$, $m_{t,k}^{j}$ defines then the density at time $t$ of individual
investors who are confident in the market signal or trend ($m\rightarrow1$),
while $1-$ $m_{t,k}^{j}$ defines the density at time $t$ of investors who
distrust that market signal ($m\rightarrow0$). The initial value
$m_{t,k=0}^{j}$ can be exogenous or endogenous to the model setting. In
particular, it can depend on $F$.

On this basis, at each time $t$, we assume that individual investors interact
within each group $j$ by subgroups of a given size for $k$ sub-periods, in
order to generate the group opinion for time $t$. In particular, for groups of
size $3$, the density after $k$ successive updates is

\[
m_{t,k}^{j}=(m_{t,k-1}^{j})^{3}+3(m_{t,k-1}^{j})^{2}(1-m_{t,k-1}^{j})\text{,}
\]

where $m_{t,k-1}^{j}$ is the proportion of agents who are confident in the
market signal at a distance of $(k-1)$ updates from the initial time $t,k=0$.
For groups of size $4$, the density after $n$ successive updates is

\[
m_{t,k}^{j}=(m_{t,k-1}^{j})^{4}+4(m_{t,k-1}^{j})^{3}\{1-m_{t,k-1}
^{j}\}+6(1-p)(m_{t,k-1}^{j})^{2}\{1-m_{t,k-1}^{j})^{2},
\]

where $m_{t,k-1}^{j}$ is the proportion of agents who are confident in the
market signal at a distance of $(k-1)$ updates from the initial time $t,k=0$.
The last term includes the tie case contribution (where two "believers"
confronted with two "distrusters") weighted with the probability $p$. Then,
the social mood (density) goes down to $0$ with probability $(1-p)$ and up to
$1$\ with probability $p$. For a mixture of group sizes with the probability
distribution $a_{i}$ with the constraint $\sum_{i=1}^{L}a_{i}=1$ - where $L$
is the largest group size and $i$ refers to the group size:

\begin{align}
m_{t,k}^{j}  &  =\sum_{i=1}^{L}a_{i}\{\sum_{j=N[\frac{i}{2}+1]}^{i}C_{j}
^{i}(m_{t,k-1}^{j})^{j}(1-m_{t,k-1}^{j})^{(i-j)}+\nonumber\\
&  (1-p)V(i)C_{\frac{i}{2}}^{i}(m_{t,k-1}^{j})^{\frac{i}{2}}
(1-m_{t,k-1}^{j})^{\frac{i}{2}}\}\text{,}
\end{align}
where $C_{j}^{i}\equiv\frac{i!}{(i-j)!j!}$, $N[\frac{i}{2}+1]\equiv$ Integer
Part of $(\frac{i}{2}+1)$, $m_{t,k-1}^{j}$ is the proportion of agents who
believe in the market signal after $(k-1)$ updates, and $V(i)\equiv{N[\frac
{i}{2}]-N[\frac{i-1}{2}]}$. This implies $V(i)=1$ for $i$ even and $V(i)=0$
for $i$ odd. The
proportion of "distrusters" is then $1-m_{t,k}^{j}$.\bigskip\bigskip

It is worth to emphasize that the Galam model of opinion dynamics tangles up three main mechanisms to produce a threshold opinion dynamics among two competing choices within an ensemble of investors. The first mechanism is exogenous and combines all effects which act directly and individually on the agent to influence its own personal choice, here to trust or distrust the current trend of the market. It determines the initial share $m_{t,k=0}^{j}$  of investors who are respectively confident to or distrusting the market price trend. The two other mechanisms are endogenous to the ensemble of interacting investors. 

One mechanism embeds a social mimetic effect using a local majority rule, i.e., agents confront their actual choice with the ones of a small group of other agents and update their respective choices following the choice which was locally majority within the group. At the collective global level, this interactive process produces a threshold dynamics  for which the tipping point is located at precisely fifty percent: The choice which starts with an initial support of more than fifty percent of the investors will drive the market along its direction.

The second mechanism is more subtle and depends on the occurrence of a local doubt within a group of investors which are settling their respective opinions. In such a case, all the involved agents converge to just one common belief about the market price trend and adopt it as their own choice. Accordingly, within an ensemble of investors, with $m_{t,k}^{j}$ percent of them expecting the trend to be positive, a local doubting group of even size may decide to either trust the trend with a probability of $p$ and distrust it with a probability of $(1-p)$.

The breaking contribution of the leading common belief is to unbalance drastically the threshold dynamics by placing the tipping point at a value which can be as low as $15 \%$ for the choice which goes along the common belief, and as high as  $85 \%$ for the choice which contradicts the common belief \cite{mino, hetero}. For the case of group of size four used in this work, we have respectively $23 \%$ and $77 \%$ for the tipping points. This second mechanism illustrates how the common belief shared by some groups of investors can shape substantially the working of the market pricing \cite{bulle}over time.

\section{The formation of the market clearing price}

The formation of the market clearing price $p_{t+1}^{\ast}$ over time depends
on the aggregation of individual bids of demand and supply at each period $t$.
In particular, every shareholder ($j=S$) $i$ wishes to sell if $p_{t+1}^{\ast
}\geq\left.  E_{t}(p_{t+1})\right\vert _{i}^{S}$, while every potential buyer
($j=D$) $i$ wishes to buy if $p_{t+1}^{\ast}\leq\left.  E_{t}(p_{t+1}
)\right\vert _{i}^{D}$. By assuming uniform distribution of individual
investors within each group $j=S,D$, the individual price expectation $\left.
E_{t}(p_{t+1})\right\vert _{i}^{j}$ of investor $i$ belonging to group $j$ can
be rewritten as a function of expectations expressed by investors $i=0$ and
$i=1$ defined as follows:

\begin{align*}
\left.  \varepsilon_{t}\right\vert _{0}^{j}  &  \equiv\left(  \left.
E_{t-1}(p_{t})\right\vert _{0}^{j}-p_{t}\right) \\
\left.  \varepsilon_{t}\right\vert _{1}^{j}  &  \equiv\left(  \left.
E_{t-1}(p_{t})\right\vert _{1}^{j}-p_{t}\right)  \text{.}
\end{align*}

Individual price expectation by investor $i$ may then be described as follows:

\[
\left.  E_{t}(p_{t+1})\right\vert _{i}^{j}=p_{t}+m_{t,k}^{j}\left(
p_{t}-p_{t-1}\right)  -\left(  \beta_{0}^{j}\left(  1-\varphi_{i}\right)
\varepsilon_{0,t}^{j}+\beta_{1}^{j}\varphi_{i}\varepsilon_{1,t}^{j}\right)
+\varphi_{i}\gamma^{j}F_{t}
\]

Aggregated demand and supply are now defined by the focal prices of four
representative agents with $i=0$ and $i=1$ $\forall j=S,D$. By defining:

\begin{align*}
\overline{P_{t}^{j}}  &  \equiv\max\arg\left[  \left.  E_{t}(p_{t+1}
)\right\vert _{i=0}^{j}\text{; }\left.  E_{t}(p_{t+1})\right\vert _{i=1}
^{j}\right] \\
\underline{P_{t}^{j}}  &  \equiv\min\arg\left[  \left.  E_{t}(p_{t+1}
)\right\vert _{i=0}^{j}\text{; }\left.  E_{t}(p_{t+1})\right\vert _{i=1}
^{j}\right]  \text{,}
\end{align*}

the aggregate functions of supply $x_{t+1}^{S}$ and demand $x_{t+1}^{D}$
integrate individual bids as follows:

\[
\left\{
\begin{array}
[c]{c}
x_{t+1}^{S}=\int_{\underline{P_{t}^{S}}}^{p_{t+1}^{\ast}}\frac{1}
{\overline{P_{t}^{S}}-\underline{P_{t}^{S}}}dx\\
x_{t+1}^{D}=\int_{p_{t+1}^{\ast}}^{\overline{P_{t}^{D}}}\frac{1}
{\overline{P_{t}^{D}}-\underline{P_{t}^{D}}}dx
\end{array}
\right.
\]

or, equivalently:

\begin{equation}
x_{t+1}^{S}=\left(
\begin{array}
[c]{c}
0\text{ \ if \ }p_{t+1}^{\ast}\leq\underline{P_{t}^{S}}\\
\frac{p_{t+1}^{\ast}-\underline{P_{t}^{S}}}{\overline{P_{t}^{S}}
-\underline{P_{t}^{S}}}\text{ \ if \ }\underline{P_{t}^{S}}<p_{t+1}^{\ast
}<\overline{P_{t}^{S}}\\
1\text{ \ if \ }p_{t+1}^{\ast}\geq\overline{P_{t}^{S}}
\end{array}
\right.
\end{equation}

\begin{equation}
x_{t+1}^{D}=\left(
\begin{array}
[c]{c}
1\text{ \ if \ }p_{t+1}^{\ast}\leq\underline{P_{t}^{D}}\\
\frac{\overline{P_{t}^{D}}-p_{t+1}^{\ast}}{\overline{P_{t}^{D}}-\underline
{P_{t}^{D}}}\text{ \ if \ }\underline{P_{t}^{D}}<p_{t+1}^{\ast}<\overline
{P_{t}^{D}}\\
0\text{ \ if \ }p_{t+1}^{\ast}\geq\overline{P_{t}^{D}}\text{.}
\end{array}
\right.
\end{equation}

The necessary condition for the existence of a market clearing price
$p_{t+1}^{\ast}$ (implying that both demand and supply are different from
zero) is

\[
\underline{P_{t}^{S}}\leq p_{t+1}^{\ast}\leq\overline{P_{t}^{D}}
\]

This condition implies two different scenarios:

\begin{enumerate}
\item if $\overline{P_{t}^{D}}\leq\underline{P_{t}^{S}}$, there is not
matching between demand and supply;\ therefore, no exchange transactions
occur, and the share Exchange does not fix any updated clearing price at
period $t$; at the next period $t+1$, investors will then observe a special
no-clearing price $p^{NC}$generated by the market-making process according to
some external rule or device;

\item if $\overline{P_{t}^{D}}>\underline{P_{t}^{S}}$, there is matching, and
the market clearing price $p^{C}$ is defined as the price that makes demand
equal to supply.\footnote{The Walrasian auction is included by this scenario
when the whole share offer is satisfied.}
\end{enumerate}

On this basis, the market clearing price at period $t$ is

\[
p_{t+1}^{\ast}=\left\{
\begin{array}
[c]{cc}
p^{NC} & \text{if }\overline{P_{t}^{D}}\leq\underline{P_{t}^{S}}\\
p^{C} & \text{if }\overline{P_{t}^{D}}>\underline{P_{t}^{S}}
\end{array}
\right.
\]

Let assume that the no-clearing price $p^{NC}$ is fixed according to the
following rule:

\[
p^{NC}=p_{t}+\epsilon
\]

where $\epsilon$ is the smallest tick value available on the share Exchange.
Furthermore, concerning the clearing price $p^{C}$, demand is equal to supply if

\[
\frac{p^{C}-\underline{P_{t}^{S}}}{\overline{P_{t}^{S}}-\underline{P_{t}^{S}}
}=\frac{\overline{P_{t}^{D}}-p^{C}}{\overline{P_{t}^{D}}-\underline{P_{t}^{D}
}}\text{, implying that}
\]

\[
p^{C}=\frac{\overline{P_{t}^{D}}\left(  \overline{P_{t}^{S}}-\underline
{P_{t}^{S}}\right)  +\underline{P_{t}^{S}}\left(  \overline{P_{t}^{D}
}-\underline{P_{t}^{D}}\right)  }{\left(  \overline{P_{t}^{D}}-\underline
{P_{t}^{D}}\right)  +\left(  \overline{P_{t}^{S}}-\underline{P_{t}^{S}
}\right)  }
\]

Therefore, the market clearing price $p_{t+1}^{\ast}$ at time $t$ is:

\begin{equation}
p_{t+1}^{\ast}=\left\{
\begin{array}
[c]{cc}
p^{NC}=p_{t}+\epsilon & \text{if }\overline{P_{t}^{D}}\leq\underline{P_{t}
^{S}}\\
p^{C}=\frac{\overline{P_{t}^{D}}\left(  \overline{P_{t}^{S}}-\underline
{P_{t}^{S}}\right)  +\underline{P_{t}^{S}}\left(  \overline{P_{t}^{D}
}-\underline{P_{t}^{D}}\right)  }{\left(  \overline{P_{t}^{D}}-\underline
{P_{t}^{D}}\right)  +\left(  \overline{P_{t}^{S}}-\underline{P_{t}^{S}
}\right)  } & \text{if }\overline{P_{t}^{D}}>\underline{P_{t}^{S}}\text{.}
\end{array}
\right.
\end{equation}
\bigskip

\section{The dynamics of the market clearing price}

In order to analyze the dynamics of the market clearing price (when it exists,
i.e., $p_{t+1}^{\ast}=p^{C}$) over time, let define $\forall j=S,D$:

\begin{align*}
\mathbf{P}^{j}\left(  n\right)   &  \equiv
{\displaystyle\sum\limits_{n=1}^{t}}
\left(  -\beta_{0}^{j}\right)  ^{n}\left(  p_{t-n}+m_{t-n}^{j}\left(
p_{t-n}-p_{t-n-1}\right)  -p_{t-n+1}\right) \\
\mathbf{F}^{j}\left(  n\right)   &  \equiv%
{\displaystyle\sum\limits_{n=0}^{t}}
\left[  \left(  -\beta_{0}^{j}\right)  ^{n}\left(  \gamma^{j}F_{t-n}\right)
\right] \\
\mathbf{L}^{j}\left(  \mathbf{P}\left(  n\right)  ,\mathbf{F}\left(  n\right)
\right)   &  \equiv\left[  \frac{\left\vert \left(  \beta_{1}^{j}-\beta
_{0}^{j}\right)  \cdot\mathbf{P}^{j}\left(  n\right)  +\mathbf{F}^{j}\left(
n\right)  \right\vert }{%
{\displaystyle\sum\limits_{j=S,D}}
\left\vert \left(  \beta_{1}^{j}-\beta_{0}^{j}\right)  \cdot\mathbf{P}
^{j}\left(  n\right)  +\mathbf{F}^{j}\left(  n\right)  \right\vert }\right]
^{-1}\\
\mathbf{M}^{j}(\mathbf{P}\left(  n\right)  ,\mathbf{F}\left(  n\right)  )  &
\equiv\left(  \beta_{1}^{j}-\beta_{0}^{j}\right)  \cdot\mathbf{P}^{j}\left(
n\right)  +\mathbf{F}^{j}\left(  n\right)  \text{.}
\end{align*}

Accordingly,

\[
\left.  E_{t}(p_{t+1})\right\vert _{i}^{j}=p_{t}+m_{t,k}^{j}\left(
p_{t}-p_{t-1}\right)  +\beta_{0}^{j}\cdot\mathbf{P}^{j}\left(  n\right)
+\varphi_{i}\cdot\mathbf{M}^{j}(\mathbf{\cdot}).
\]

The four rapresentative agents are then described as follows:
\begin{align*}
\forall j  &  =S,D\text{ with }\varphi_{i}=0\text{:}\left.  E_{t}
(p_{t+1})\right\vert _{i=0}^{j}=p_{t}+m_{t,k}^{j}\left(  p_{t}-p_{t-1}\right)
+\beta_{0}^{j}\cdot\mathbf{P}^{j}\left(  n\right)  \bigskip\\
\forall j  &  =S,D\text{ with }\varphi_{i}=1\text{:}\left.  E_{t}
(p_{t+1})\right\vert _{i=0}^{j}=p_{t}+m_{t,k}^{j}\left(  p_{t}-p_{t-1}\right)
+\beta_{1}^{j}\cdot\mathbf{P}^{j}\left(  n\right)  +\mathbf{F}^{j}\left(
n\right)  \text{.}
\end{align*}

By computation, the market clearing price function can be rewritten as follows:

\begin{align}
p_{t+1}^{\ast}  &  =p_{t}+
{\displaystyle\sum\limits_{j=S,D}}
\left[  \frac{\left(  m_{t}^{j}\left(  p_{t}-p_{t-1}\right)  +\mathbf{P}
^{j}\left(  \cdot\right)  \right)  }{\mathbf{L}^{\lnot j}\left(
\mathbf{\cdot}\right)  }\right]  +\label{clearing_price_2}\\
&  \left\{
\begin{array}
[c]{c}
\begin{tabular}
[c]{ll}
$\left[  \frac{\mathbf{M}^{D}(\mathbf{\cdot})}{\mathbf{L}^{S}\left(
\mathbf{\cdot}\right)  }\right]  $ & if $\mathbf{M}^{j}(\cdot)>0$ $\forall
j$\\
& \\
$\left[  \frac{\mathbf{M}^{S}(\mathbf{\cdot})}{\mathbf{L}^{D}\left(
\mathbf{\cdot}\right)  }\right]  $ & if $\mathbf{M}^{j}(\cdot)<0$ $\forall
j$\\
$
{\displaystyle\sum\limits_{j=S,D}}
\left[  \frac{\mathbf{M}^{j}(\mathbf{\cdot})}{\mathbf{L}^{\lnot j}\left(
\mathbf{\cdot}\right)  }\right]  $ &
\begin{tabular}
[c]{l}
if $\mathbf{M}^{D}(\cdot)>0$\\
and $\mathbf{M}^{S}(\cdot)<0$
\end{tabular}
\\
$0$ &
\begin{tabular}
[c]{l}
if $\mathbf{M}^{D}(\cdot)<0$\\
and $\mathbf{M}^{S}(\cdot)>0$
\end{tabular}
\end{tabular}
\ \ \\
\end{array}
\right. \nonumber
\end{align}

Accordingly, the pattern of market clearing price $p_{t+1}^{\ast}$ is based on
the historical price $p_{t}$ by adding two further elements.\ The first
element comprises (for both $j=S$ and $j=D$) two sub-elements:

\begin{itemize}
\item the numerator, $m_{t,k}^{j}\left(  p_{t}-p_{t-1}\right)  +\mathbf{P}
^{j}\left(  n\right)  $, is independent from signal $F_{t}$ and dependent on
the price trend $\Delta_{t}\left(  p^{\ast}\right)  $ weighted by the current
group mood $m_{t}^{j}$,\footnote{Remember that $m_{t,k}^{j}\rightarrow1$
implies full weight to this information in order to build individual price
expectations.The mood $m$ can be influenced by the fundamental signal $F$, and
is the final result of the dynamic interaction within the group $j$ for $k$
steps occurring between $t-1$ and $t$.} and its weighted past series
$\mathbf{P}^{j}\left(  n\right)  $;

\item the denominator, $\mathbf{L}^{j}\left(  \mathbf{\cdot}\right)  $,
depends on both $\mathbf{F}^{j}\left(  n\right)  $ which represents the
weighted fundamental signal trend series, and $\mathbf{P}^{j}\left(  n\right)
$ which represents the weighted market price trend series; for each group $j$,
this sub-element weights the contribution of the price trend series to the
formation of the market clearing price at time $t$.
\end{itemize}

The second element depends on both weighted past series $\mathbf{F}^{j}\left(
\cdot\right)  $ and $\mathbf{P}^{j}\left(  \cdot\right)  $. In particular, if
$\mathbf{M}^{j}(\mathbf{\cdot})$ is positive (negative) for both groups, then
this element increases (decreases) the market clearing price. Moreover, if
$\mathbf{M}^{j}(\mathbf{\cdot})$ is negative for shareholders ($j=S$) while it
is positive for potential investors ($j=D$), then the divergence between
groups is mutually balanced on the marketplace. On the contrary, if
$\mathbf{M}^{j}(\mathbf{\cdot})$ is positive for shareholders while negative
for potential investors, then the divergence makes the whole element equal to zero.

In sum, the formation of share prices over time depends respectively on the
dynamics of the fundamental signal $F$ from one side, and the dynamics of the
clearing market price $p$ from another side. Both dynamics are shaped by the
ongoing evolution of individual and group opinions (and related bids) captured
by the structure of the model.

\section{An illustrative analysis of a particular specification of the model}

This section shall illustrate the theoretical contribution of our model by
visualizing some particular cases.\ For this purpose,let assume that
$\beta_{i}^{j}=\beta$ and $\gamma^{j}=\gamma$ $\forall j=S,D$ and $\forall i$.
This specification implies that group heterogeneity is captured by the group
mood $m_{t,k}^{j}$ and leads to the following proposition:

\begin{proposition}
If $\beta_{i}^{j}=\beta$ and $\gamma^{j}=\gamma$ $\forall j=S,D$ and $\forall
i$, the group mood $m_{t,k}^{j}$ subsumes all the group heterogeneities
between demand and supply; then, there exists only one market clearing price
case instead of the four cases defined above.
\end{proposition}

In particular, the individual price expectation function becomes:

\begin{align*}
\left.  E_{t}(p_{t+1})\right\vert _{i}^{j}  &  =p_{t}+m_{t,k}^{j}\left(
p_{t}-p_{t-1}\right)  +\\
&
{\displaystyle\sum\limits_{n=1}^{t}}
\left[  \left(  -\beta\right)  ^{n}\left(  p_{t-n}-m_{t-n}^{j}\left(
p_{t-n}-p_{t-n-1}\right)  -p_{t-n+1}\right)  \right]  +\\
&  \varphi_{i}%
{\displaystyle\sum\limits_{n=0}^{t}}
\left[  \left(  -\beta\right)  ^{n}\left(  \gamma F_{t-n}\right)  \right] \\
\text{or }\left.  E_{t}(p_{t+1})\right\vert _{i}^{j}  &  =p_{t}+m_{t,k}
^{j}\left(  p_{t}-p_{t-1}\right)  +\mathbf{P}^{j}\left(  n\right)
+\varphi_{i}\gamma^{j}\mathbf{F}^{j}\left(  n\right)  \text{.}
\end{align*}

Concerning the formation of the market clearing price, for $\beta_{i}
^{j}=\beta$ $\forall i,j$, $\mathbf{L}^{j}\left(  \mathbf{\cdot}\right)
=2$.\ Therefore, closer are $\beta_{i}^{j}$ $\forall i,j$, closer is
$\mathbf{L}^{j}\left(  \mathbf{\cdot}\right)  $ to $2$, implying that the
whole first element of equation \ref{clearing_price_2} tends to become
independent from the fundamental signal series $\mathbf{F}^{j}\left(
n\right)  $. Furthermore, when $\beta_{1}^{j},\beta_{0}^{j}=\beta^{j}$,
$\mathbf{M}^{j}(\cdot)=\mathbf{F}^{j}\left(  n\right)  $ $\forall j$:\ Closer
are $\beta_{0}^{j}$ and $\beta_{1}^{j}$ $\forall j$, closer is $\mathbf{M}
^{j}(\cdot)$ to $\mathbf{F}^{j}\left(  n\right)  $ that is independent from
the market price trend series $\mathbf{P}^{j}\left(  n\right)  $. Therefore,
this specification clearly distinguishes the dual structure of the market
clearing price dynamics which is driven by two distinct factors: the market
signal or trend $\Delta_{t}\left(  p^{\ast}\right)  $ weighted by the
evolution of groups' market sentiments, and the fundamental signal $F$. The
market clearing price becomes:

\[
p_{t+1}^{\ast}=p_{t}+
\frac12
{\displaystyle\sum\limits_{j=S,D}}
\left(  m_{t,k}^{j}\left(  p_{t}-p_{t-1}\right)  +\mathbf{P}^{j}\left(
n\right)  \right)  +\left[  \frac{\mathbf{F}\left(  n\right)  }{2}\right]
\]

where

\begin{align*}
\mathbf{P}^{j}\left(  n\right)   &  \equiv
{\displaystyle\sum\limits_{n=1}^{t}}
\left(  -\beta\right)  ^{n}\left(  p_{t-n}+m_{t-n}^{j}\left(  p_{t-n}
-p_{t-n-1}\right)  -p_{t-n+1}\right) \\
\mathbf{F}\left(  n\right)   &  \equiv
{\displaystyle\sum\limits_{n=0}^{t}}
\left[  \left(  -\beta\right)  ^{n}\left(  \gamma F_{t-n}\right)  \right]
\end{align*}

Accordingly, the dynamics of the market clearing price (when it exists) is
denoted as follows:

\[
\Delta_{t+1}\left(  p^{\ast}\right)  \equiv p_{t+1}-p_{t}=f\left(  \Delta
_{t}\left(  p^{\ast}\right)  ,m_{t,k}^{j}\right)  +g\left(  \mathbf{F}\left(
n\right)  \right)
\]

This price pattern comprises two different elements. The first element,
$f\left(  \Delta_{t}\left(  p^{\ast}\right)  ,m_{t,k}^{j}\right)  $, is a
group factor that depends on the market signal $\Delta_{t}\left(  p^{\ast
}\right)  $ weighted by the group mood $m_{t,k}^{j}$ that is collectively
assigned to the market price trend by group $j$ at time $t$. The second
element, $g\left(  \mathbf{F}\left(  n\right)  \right)  =\frac{\mathbf{F}
\left(  n\right)  }{2}$, depends on the weighted trend of the fundamental
signal $F_{t}$, with $\mathbf{F}^{j}\left(  n\right)  =\mathbf{F}^{D}\left(
n\right)  =\mathbf{F}^{S}\left(  n\right)  $ in this particular specification.
Consequently, if $\mathbf{F}\left(  n\right)  $ is positive (negative), then
$g\left(  \mathbf{F}\left(  n\right)  \right)  $ proportionally increases
(decreases) the market clearing price at time $t$.

\subsection{Illustrative case of a constant trend in the fundamental signal}

Let illustrate this particular specification of the model when the fundamental
signal experiences an alternate positive and negative trend: $F_{t}=\pm0.1$
every $10$ periods of $t$. Let assume: $\beta_{i}^{j}=\beta=0.5$; $\gamma
^{j}=\gamma=1$; $\epsilon=0.01$; $p_{0}=10$; $F_{0}=0$; $m_{t,k=0}^{S}=0.6$;
$m_{t,k=0}^{D}=0.4$; $\left.  \varepsilon_{t=0}\right\vert _{i}^{j}
=0.1\cdot(U(0;1)-U(0;1))$, $\forall j=S,D$ and $\forall i$. The group
interaction is based on groups of size $3$, $4$ (with $p^{D}=p^{S}=0.3$) and
$4$ (with $p^{D}=0.3$ and $p^{S}=0.3$).\ The probability $p$ means here that,
in case of group inderterminacy, the group belief tends to trust the market
($p\rightarrow1$) or not ($p\rightarrow0$).\ We perform simulations for
periods from $t=1$ to $t=100$, with various $k$ from\ $0$\ to $7$. In this
case, the market clearing price and the fundamental signal change at the same
rhythm $t$, while the market sentiment changes at its rhythm $k$ from $0$ (no
steps, implying no change from the initial value) to $7$\ (seven steps between
$t-1$ and $t$). Figures 1a,b,c illustrate the result.

\begin{figure}
\centering
\includegraphics[width=.7\textwidth]{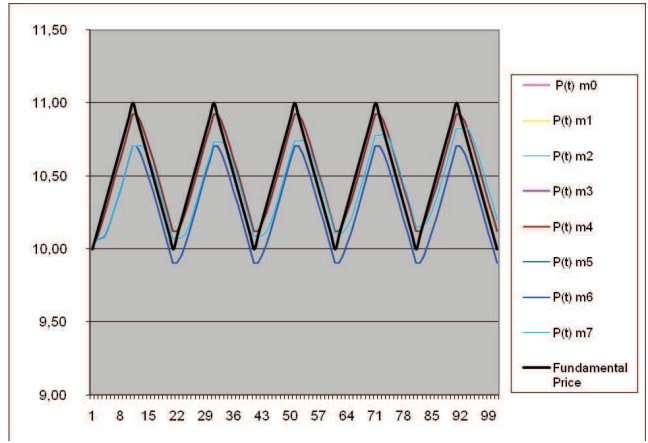}
\caption{Simulation for groups of size 3}
\label{f1}
\end{figure}

\begin{figure}
\centering
\includegraphics[width=.7\textwidth]{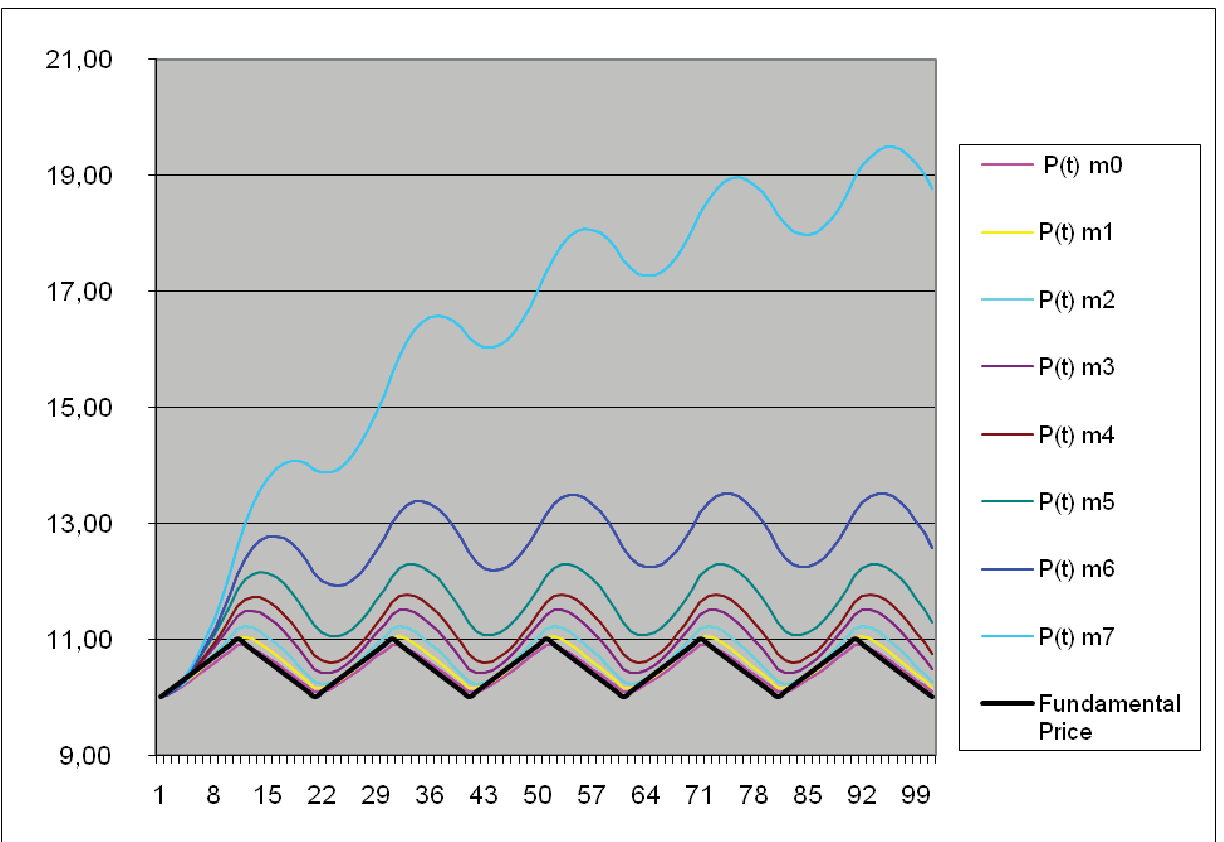}
\caption{Simulation for groups of size 4 ($p^{D}=p^{S}=0.3$}
\label{f2}
\end{figure}

\begin{figure}
\centering
\includegraphics[width=.7\textwidth]{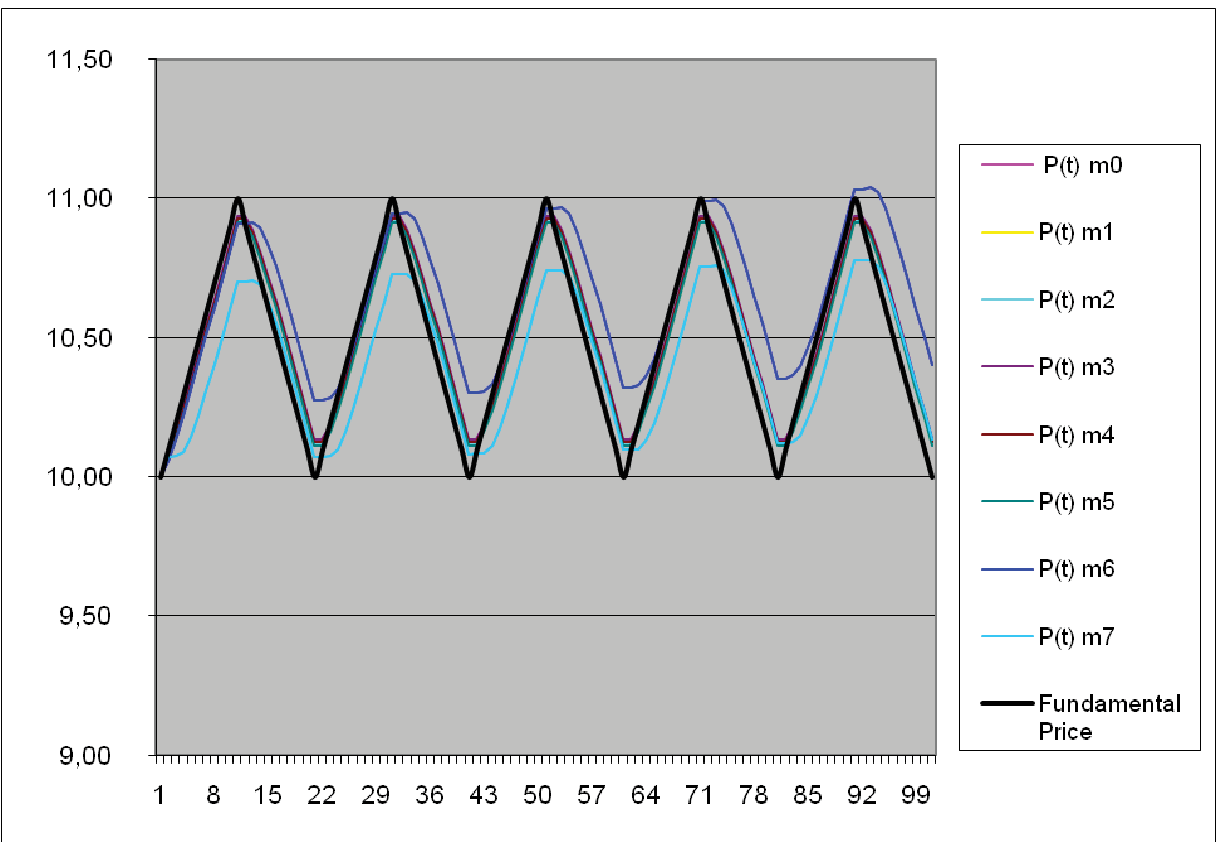}
\caption{Simulation for groups of size $4$ (with $p^{D}=0.6$ and $p^{S}=0.4$)}
\label{f3}
\end{figure}

The simulation shows that changes in the market sentiment exacerbates the
impact of the fundamental signal on the market clearing price over time. The
resulting market price remains under- or over-valuated the shares relative to
their fundamental price computed as follows:\footnote{By definition, the first
market price and the first fundamental price are equal.}

\[
p_{t}^{F}=p_{t-1}^{F}+F_{t-1}=p_{_{0}}+
{\displaystyle\sum\limits_{n=0}^{t-1}}
F_{n}\text{.}
\]

\subsection{Illustrative case of market exuberance}

Under the same conditions, let assume that the fundamental signal experiences
a random pattern: $F_{t}=U(0;1)-U(0;1)$ $\forall t\geq1$. We perform
simulations for periods from $t=1$ to $t=100$, with various $k$ from\ $0$\ to
$7$. In this case, the market clearing price and the fundamental signal change
at the same rhythm $t$, while the market sentiment changes at its rhythm $k$
from $0$ (no change) to $7$\ (seven steps between $t-1$ and $t$). Figures
2a,b,c illustrate the result.

\begin{figure}
\centering
\includegraphics[width=.7\textwidth]{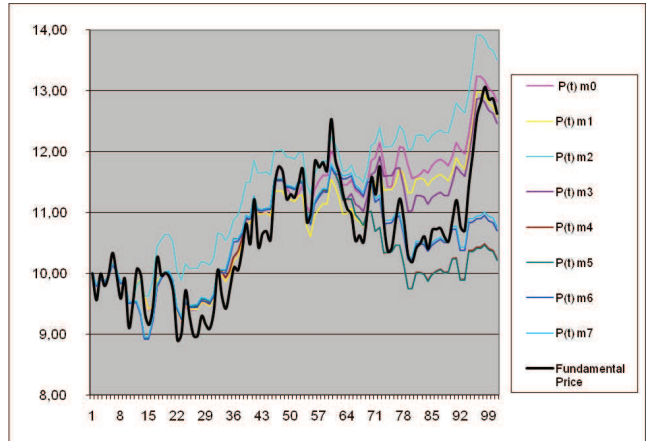}
\caption{Simulation for groups of size $3$}
\label{f4}
\end{figure}

\begin{figure}
\centering
\includegraphics[width=.7\textwidth]{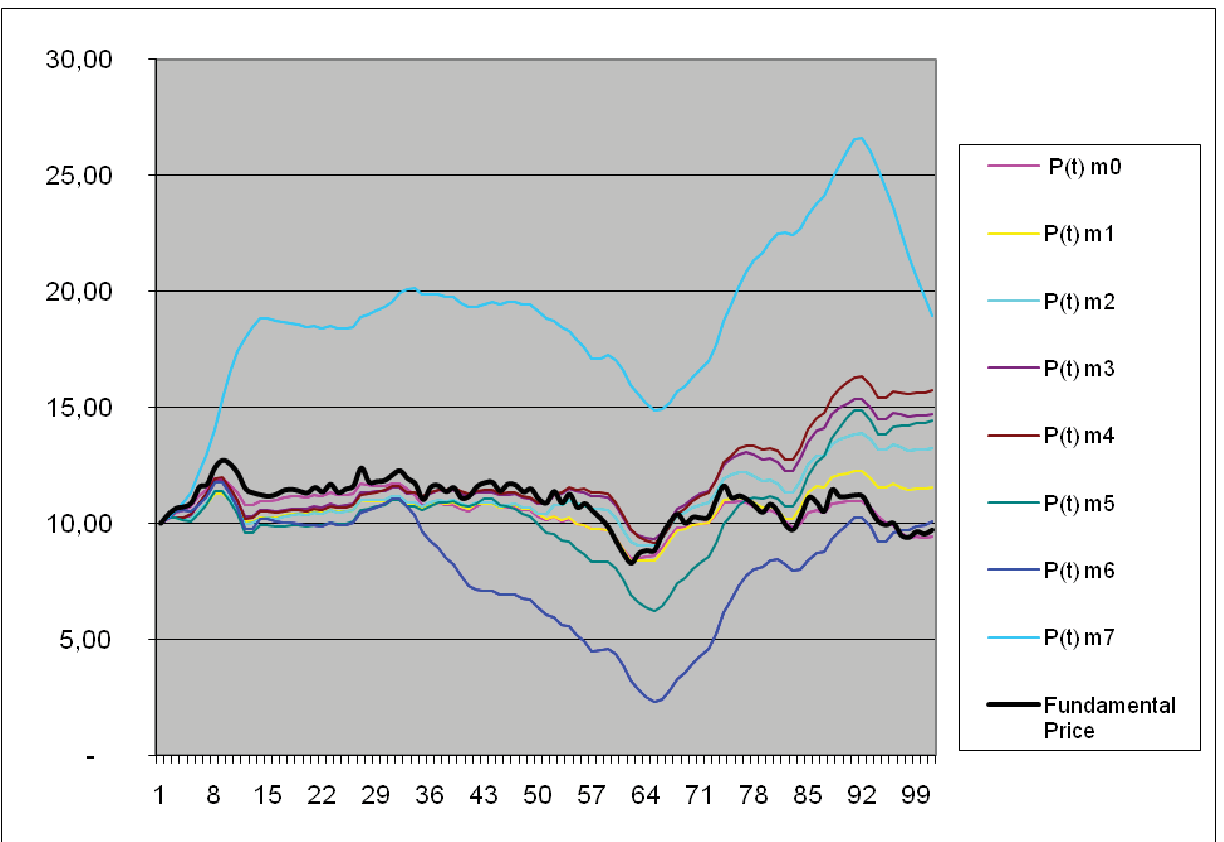}
\caption{Simulation for groups of size $4$ ($p^{D}=p^{S}=0.3$)}
\label{f5}
\end{figure}

\begin{figure}
\centering
\includegraphics[width=.7\textwidth]{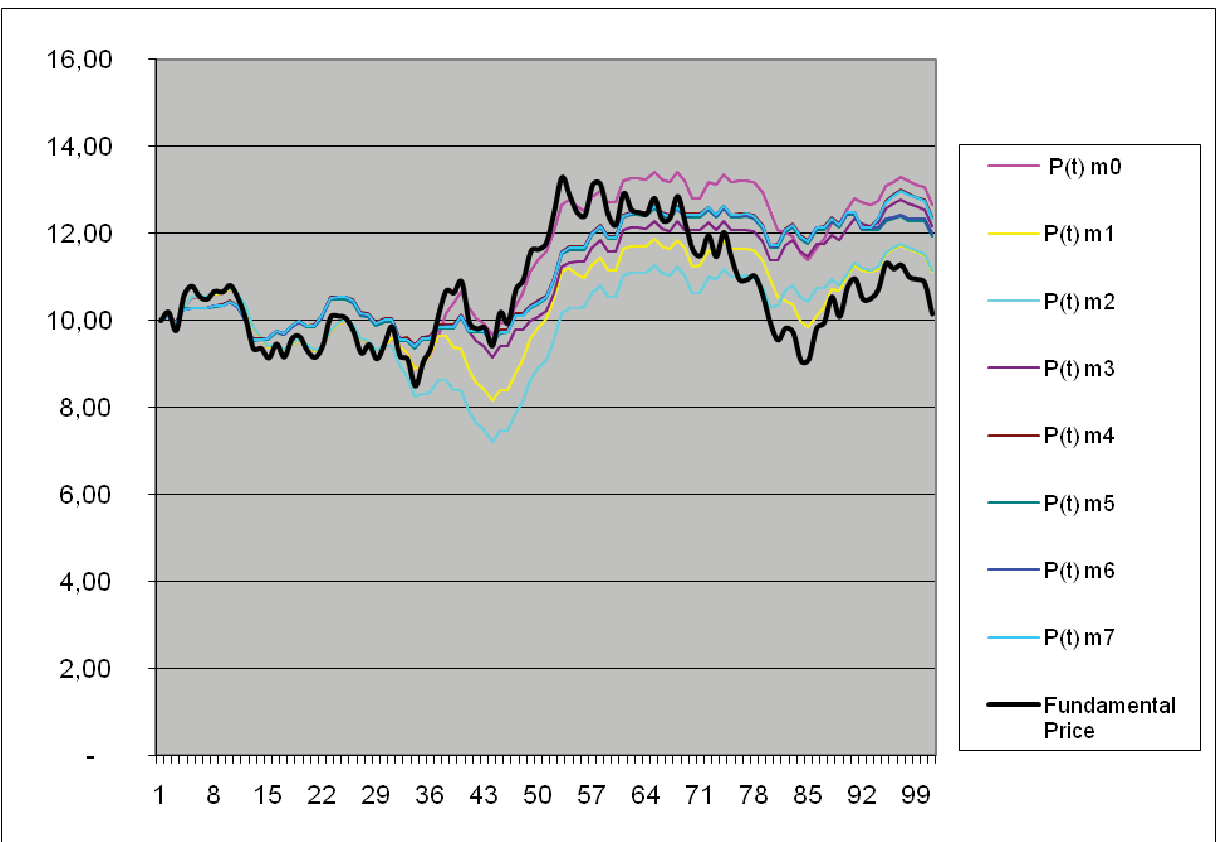}
\caption{Simulation for groups of size $4$ (with $p^{D}=0.6$ and $p^{S}=0.4$)
}
\label{f6}
\end{figure}

The simulation shows that changes in the market sentiment exacerbates the
market exuberance around the path provided by the fundamental price $p_{t}
^{F}$. This result is in line with theoretical and empirical analyses of
market exuberance discussed by  \cite{Shiller} and  \cite{LeRoy}  among others.

\subsection{Illustrative case of market disconnection from fundamental price}

The latter case shows a distinctive pattern where the market price disconnects
from the fundamental price over time.\ Figures 3a,b,c illustrate this result.

\begin{figure}
\centering
\includegraphics[width=.7\textwidth]{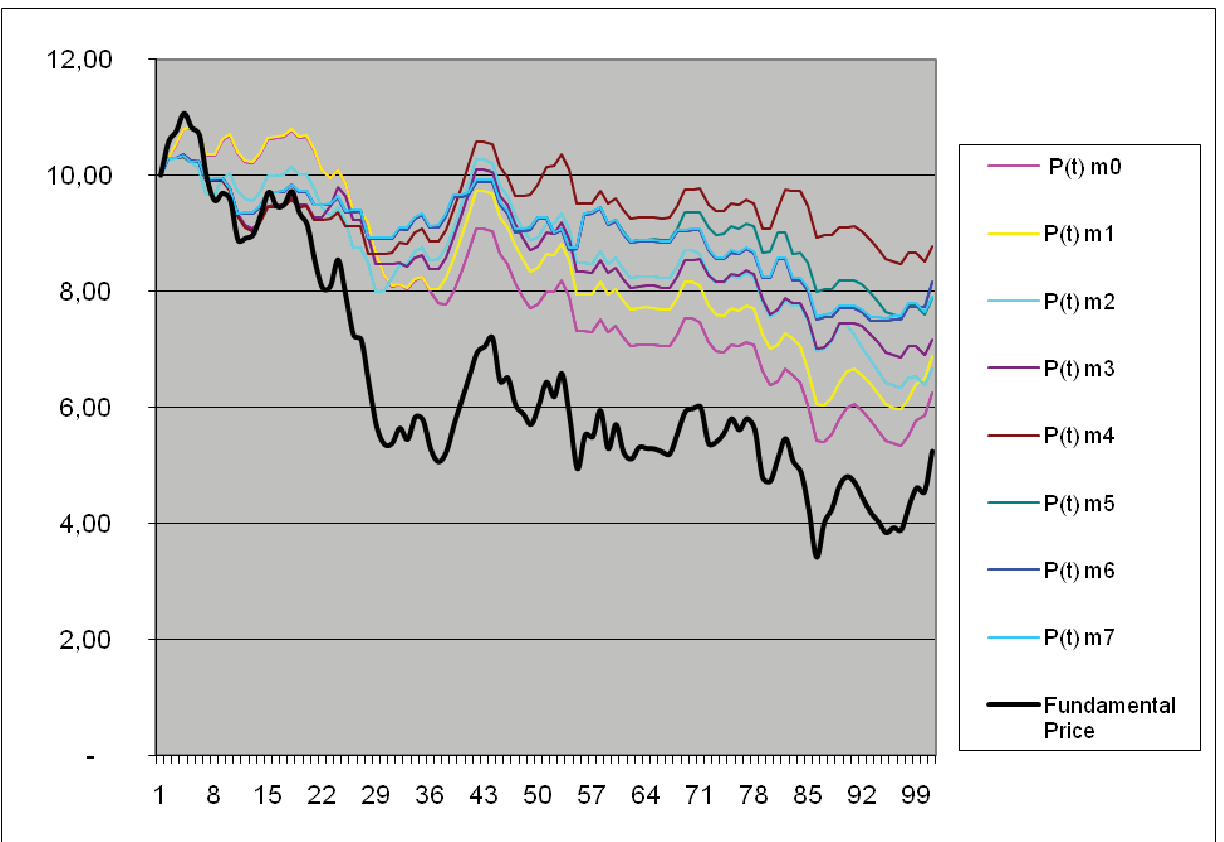}
\caption{Simulation for groups of size $3$}
\label{f7}
\end{figure}

\begin{figure}
\centering
\includegraphics[width=.7\textwidth]{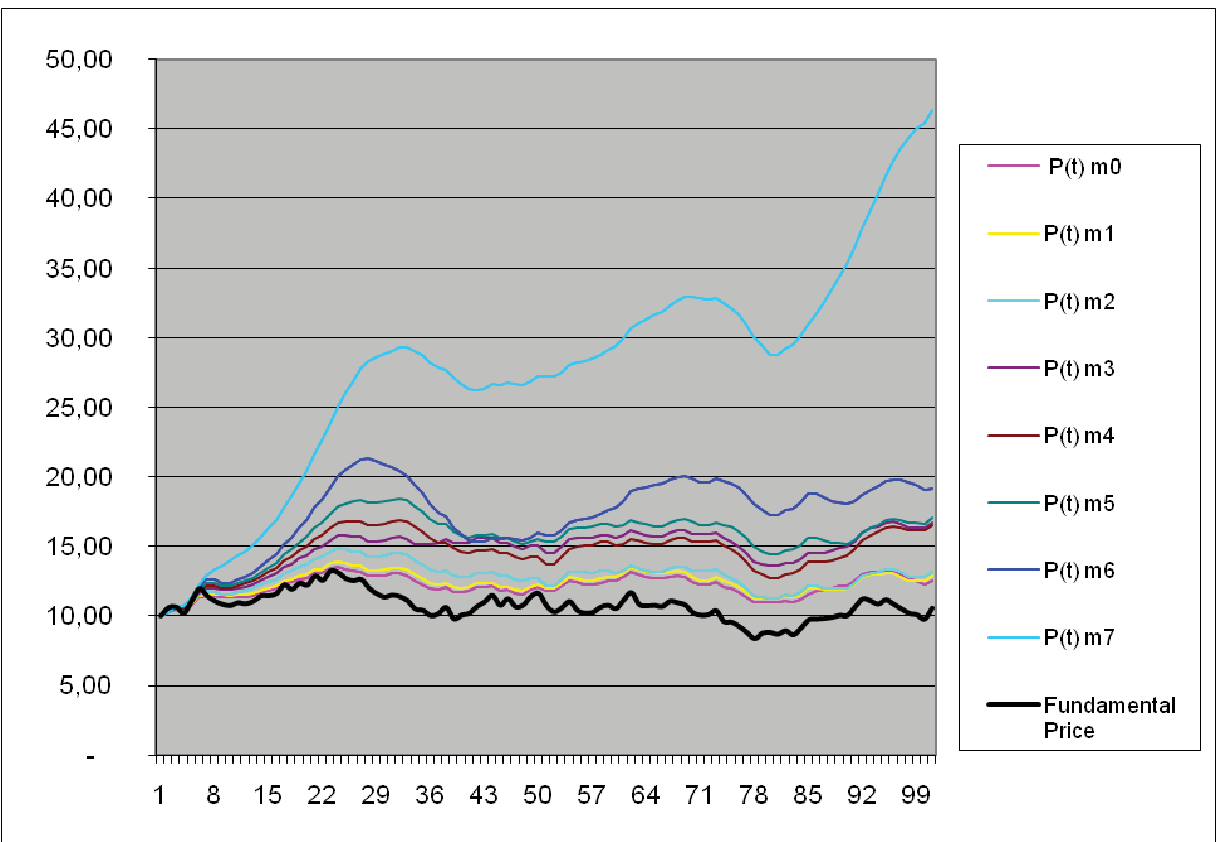}
\caption{Simulation for groups of size $4$ ($p^{D}=p^{S}=0.3$)}
\label{f8}
\end{figure}

\begin{figure}
\centering
\includegraphics[width=.7\textwidth]{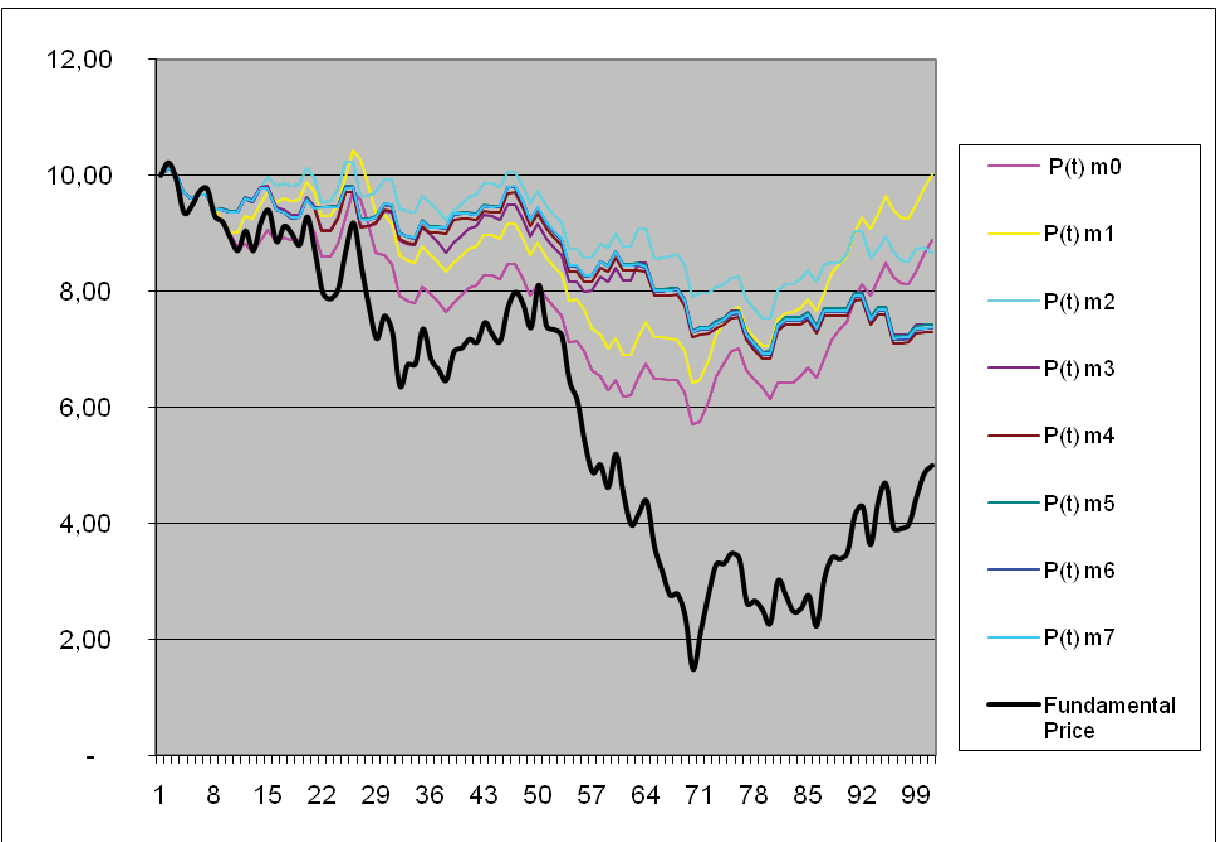}
\caption{Simulation for groups of size $4$ (with $p^{D}=0.6$ and $p^{S}=0.4$)}
\label{f9}
\end{figure}

The simulation shows that the dynamics of the clearing price may be
disconnected by the dynamics of fundamental price $p_{t}^{F}$ over time.\ This
implies that the formation of a market clearing price over time is not
sufficient to assure that market pricing is aligned on the fundamental price
that arises from fundamentals that are common knowledge among heterogeneous
market participants.

\section{Conclusive remarks}

Financial economic models often assume that investors know (or agree on) the
fundamental value of the firm's shares that are traded, easing the passage
from the individual to the collective dimension of the financial system
generated by the Share Exchange over time.\ Our model relaxes that heroic
assumption of one unique "true value" and deals with the formation of share
market prices through the dynamic formation of individual and social opinions
(or beliefs) based upon a fundamental signal of economic performance and
position of the firm, the forecast revision by heterogeneous individual
investors, and their social mood or sentiment about the ongoing state of the
market pricing process. Market clearing price formation is then featured by
individual and group dynamics that make its collective dimension irreducible
to its individual level.

This dynamic holistic approach provides a better understanding of the market
exuberance generated by the Share Exchange over time. This exuberance depends
not only on individual biases or mistakes, but also on dynamic and collective
dimensions that arise from the interaction of individuals among them and with
evolving collective structures over time. Our model captures this collective
dimension through the evolution of available common knowledge on the economic
performance and position of the firm (fundamental or firm-specific
information), as well as through the evolving social mood or sentiment on the
current state of the market (or the industry, or the whole economy). While the
former can be related to information release by accounting reporting and
disclosure, the latter can be related to investors' confidence and financial
analysts' consensus, and their respective evolution over socio-economic time
and space where the financial market is embedded.



\newpage

\section*{Appendix}

For sake of completeness, we provide here various other versions of the market
price equation.\ Starting from equation \ref{clearing_price_2}, let define:

\begin{align*}
\lambda_{t}  &  \equiv\frac{\left\vert \mathbf{M}^{S}\right\vert }{\left\vert
\mathbf{M}^{S}\right\vert +\left\vert \mathbf{M}^{D}\right\vert }
=\mathbf{L}^{S}\left(  \mathbf{P}\left(  n\right)  ,\mathbf{F}\left(
n\right)  \right)  ^{-1}\\
\left(  1-\lambda_{t}\right)   &  \equiv\frac{\left\vert \mathbf{M}
^{D}\right\vert }{\left\vert \mathbf{M}^{S}\right\vert +\left\vert
\mathbf{M}^{D}\right\vert }=\mathbf{L}^{D}\left(  \mathbf{P}\left(  n\right)
,\mathbf{F}\left(  n\right)  \right)  ^{-1}\\
\text{ with }\mathbf{M}^{j}(\mathbf{\cdot})  &  =\left(  \beta_{1}^{j}
-\beta_{0}^{j}\right)  \cdot\mathbf{P}^{j}\left(  n\right)  +\mathbf{F}
^{j}\left(  n\right)  =-\left(  \beta_{1}^{j}\varepsilon_{1,t}^{j}-\beta
_{0}^{j}\varepsilon_{0,t}^{j}\right)  +\gamma^{j}F_{t}
\end{align*}
Or, equivalently:

\begin{align*}
\lambda_{t}  &  =\frac{\left\vert
\begin{array}
[c]{c}
\left(  \beta_{1}^{S}-\beta_{0}^{S}\right)
{\displaystyle\sum\limits_{n=1}^{t}}
\left[  \left(  -\beta_{0}^{S}\right)  ^{n}\left(  p_{t-n}+m_{t-n}^{S}\left(
p_{t-n}-p_{t-n-1}\right)  -p_{t-n+1}\right)  \right]  +\\
{\displaystyle\sum\limits_{n=0}^{t}}
\left[  \left(  -\beta_{0}^{D}\right)  ^{n}\left(  \gamma^{D}F_{t-n}\right)
\right]
\end{array}
\right\vert }{
{\displaystyle\sum\limits_{j=S,D}}
\left\vert
\begin{array}
[c]{c}
\left(  \beta_{1}^{j}-\beta_{0}^{j}\right)
{\displaystyle\sum\limits_{n=1}^{t}}
\left[  \left(  -\beta_{0}^{j}\right)  ^{n}\left(  p_{t-n}+m_{t-n}^{j}\left(
p_{t-n}-p_{t-n-1}\right)  -p_{t-n+1}\right)  \right]  +\\
{\displaystyle\sum\limits_{n=0}^{t}}
\left[  \left(  -\beta_{0}^{j}\right)  ^{n}\left(  \gamma^{j}F_{t-n}\right)
\right]
\end{array}
\right\vert }\\
&  \text{and}\\
\left(  1-\lambda_{t}\right)   &  =\frac{\left\vert
\begin{array}
[c]{c}
\left(  \beta_{1}^{D}-\beta_{0}^{D}\right)
{\displaystyle\sum\limits_{n=1}^{t}}
\left[  \left(  -\beta_{0}^{D}\right)  ^{n}\left(  p_{t-n}+m_{t-n}^{D}\left(
p_{t-n}-p_{t-n-1}\right)  -p_{t-n+1}\right)  \right]  +\\
{\displaystyle\sum\limits_{n=0}^{t}}
\left[  \left(  -\beta_{0}^{D}\right)  ^{n}\left(  \gamma^{D}F_{t-n}\right)
\right]
\end{array}
\right\vert }{
{\displaystyle\sum\limits_{j=S,D}}
\left\vert
\begin{array}
[c]{c}
\left(  \beta_{1}^{j}-\beta_{0}^{j}\right)
{\displaystyle\sum\limits_{n=1}^{t}}
\left[  \left(  -\beta_{0}^{j}\right)  ^{n}\left(  p_{t-n}+m_{t-n}^{j}\left(
p_{t-n}-p_{t-n-1}\right)  -p_{t-n+1}\right)  \right] \\
+
{\displaystyle\sum\limits_{n=0}^{t}}
\left[  \left(  -\beta_{0}^{j}\right)  ^{n}\left(  \gamma^{j}F_{t-n}\right)
\right]
\end{array}
\right\vert }
\end{align*}

Therefore, the market clearing equation can be rewritten as follows:

\begin{align*}
p_{t+1}^{\ast}  &  =p_{t}+\lambda_{t}m_{t}^{D}\left(  p_{t}-p_{t-1}\right)
+\left(  1-\lambda_{t}\right)  m_{t}^{S}\left(  p_{t}-p_{t-1}\right)  +\\
&  \lambda_{t}\left(  \beta_{0}^{D}\varepsilon_{0,t}^{D}\right)  +\left(
1-\lambda_{t}\right)  \left(  \beta_{0}^{S}\varepsilon_{0,t}^{S}\right)  +\\
&  \left\{
\begin{tabular}
[c]{ll}
$\lambda_{t}\left[  \left(  \beta_{1}^{D}\varepsilon_{1,t}^{D}-\beta_{0}
^{D}\varepsilon_{0,t}^{D}\right)  +\gamma_{t}^{D}F_{t}\right]  $ & $\text{if
}\mathbf{M}^{j}>0$\\
$\left(  1-\lambda_{t}\right)  \left[  \left(  \beta_{1}^{S}\varepsilon
_{1,t}^{S}-\beta_{0}^{S}\varepsilon_{1,t}^{S}\right)  +\gamma_{t}^{S}
F_{t}\right]  $ & if $\mathbf{M}^{j}<0$\\
$
\begin{array}
[c]{c}
\lambda_{t}\left[  \left(  \beta_{1}^{D}\varepsilon_{1,t}^{D}-\beta_{0}
^{D}\varepsilon_{0,t}^{D}\right)  +\gamma_{t}^{D}F_{t}\right]  +\\
\left(  1-\lambda_{t}\right)  \left[  \left(  \beta_{1}^{S}\varepsilon
_{1,t}^{S}-\beta_{0}^{S}\varepsilon_{1,t}^{S}\right)  +\gamma_{t}^{S}
F_{t}\right]
\end{array}
$ &
\begin{tabular}
[c]{l}
if $\mathbf{M}^{D}>0$\\
and $\mathbf{M}^{S}<0$
\end{tabular}
\\
$0$ &
\begin{tabular}
[c]{l}
if $\mathbf{M}^{D}<0$\\
and $\mathbf{M}^{S}>0$
\end{tabular}
\end{tabular}
\ \ \right.
\end{align*}

Or, equivalently:

\begin{align*}
p_{t+1}^{\ast}  &  =p_{t}+\lambda_{t}m_{t}^{D}\left(  p_{t}-p_{t-1}\right)
+\left(  1-\lambda_{t}\right)  m_{t}^{S}\left(  p_{t}-p_{t-1}\right)  +\\
&  \lambda_{t}
{\displaystyle\sum\limits_{n=1}^{t}}
\left[  \left(  -\beta_{0}^{D}\right)  ^{n}\left(  p_{t-n}+m_{t-n}^{D}\left(
p_{t-n}-p_{t-n-1}\right)  -p_{t-n+1}\right)  \right]  +\\
&  \left(  1-\lambda_{t}\right)
{\displaystyle\sum\limits_{n=1}^{t}}
\left[  \left(  -\beta_{0}^{S}\right)  ^{n}\left(  p_{t-n}+m_{t-n}^{S}\left(
p_{t-n}-p_{t-n-1}\right)  -p_{t-n+1}\right)  \right]  +\\
&  \left\{  \
\begin{tabular}
[c]{ll}
$\lambda_{t}\left[  \mathbf{M}^{D}(\mathbf{\cdot})\right]  $ & if
$\mathbf{M}^{j}>0$ $\forall j$\\
$\left(  1-\lambda_{t}\right)  \left[  \mathbf{M}^{S}(\mathbf{\cdot})\right]
$ & if $\mathbf{M}^{j}<0$ $\forall j$\\
$\lambda_{t}\left[  \mathbf{M}^{D}(\mathbf{\cdot})\right]  +\left(
1-\lambda_{t}\right)  \left[  \mathbf{M}^{S}(\mathbf{\cdot})\right]  $ &
\begin{tabular}
[c]{l}
if $\mathbf{M}^{D}>0$\\
and $\mathbf{M}^{S}<0$
\end{tabular}
\\
$0$ &
\begin{tabular}
[c]{l}
if $\mathbf{M}^{D}<0$\\
and $\mathbf{M}^{S}>0$
\end{tabular}
\end{tabular}
\ \right.
\end{align*}


\begin{thebibliography}{12}

\bibitem {hom} C. H. Hommes, Heterogeneous Agent Models in Economics and Finance, In Leigh Tesfatsion and Kenneth L. Judd (ed.), Handbook of Computational Economics, chapter 23,  1109-1186 (2006)

\bibitem {cris} M. Cristelli, L. Pietronero, A. Zaccaria, Critical Overview of Agent-based Model for Economics, arXiv:1101.1847v1 [q-fin.TR]  (2011)

\bibitem {chakra} A. Chakraborti., I. M.Toke, M. Patriarca and F. Abergel, e-Print arXiv:0909.1974, (2009)

\bibitem {sama} E. Samanidou., E. Zschischang, D. Stauffer and T. Lux., Report on Progress in Physics, 70 (2007)

\bibitem {fortunato-review} C. Castellano, S. Fortunato and V. Loreto, ``Statistical physics of social dynamics", Rev. Mod. Phys. 81, 591-646 (2009)

\bibitem {galam-review} S. Galam, ``Sociophysics: a review of Galam models", Int. J. Mod. Phys. C \textbf{19},  409-440 (2008)

\bibitem {bulle} S. Galam, ``Valeurs fondamentales et croyances collectives", in  \textit{Critique de la valeur fondamentale}, Chap. 5, 99-115, Springer (Paris), C. Walter et E. Brian (Eds), (2008)

\bibitem {martins} A. C. R. Martins, C. B. Pereira and R. Vicente, ``An Opinion Dynamics Model for the Diffusion of Innovations", Physica. A \textbf{388}, 3225-3232 (2009)

\bibitem{ausloos-religion} M. Ausloos and  F. Petroni , ``Statistical dynamics of religions and adherents", Euro. Phys. Lett. \textbf{77},  38002 (1-4), (2007)

\bibitem {pierluigi}  P. Contucci and S. Ghirlanda, ``Modeling Society with Statistical Mechanics: an Application to Cultural Contact and Immigration", Quality and Quantity  \textbf{41}, 569-578 (2007) 

\bibitem {schneider} J. J. Schneider and C. Hirtreiter, ``The Impact of election results on the member numbers of the large parties in bavaria and germany, Int. J. Mod. Phys. C \textbf{16},  1165-1215 (2005)

\bibitem{sznajd} K. Sznajd-Weron and J. Sznajd,``Opinion evolution in closed community",  Int. J. Mod. Phys. C \textbf{11}, 1157-1165 (2000)

\bibitem {deffuant} G. Deffuant, D. Neau, F. Amblard and G. Weisbuch, ``Mixing beliefs among interacting agents", Advances in Complex Systems \textbf{3}, 87-98 (2000)

\bibitem {pair} S. Galam, ``Collective beliefs versus individual inflexibility: The unavoidable biases
of a public debate", Physica A, in press  (2011)

\bibitem {mino} S. Galam, 
``Minority Opinion Spreading in Random Geometry", Eur. Phys. J. B \textbf{25} Rapid Note, 403-406 (2002) 

\bibitem {hetero} S. Galam, ``Heterogeneous beliefs, segregation, and extremism in the making of public opinions", Phys. Rev. E \textbf{71}, 046123 (2005)

\bibitem {Shubik} M. Shubik, \textquotedblleft Accounting and its Relationship to
General Equilibrium Theory\textquotedblright\ , reprinted in Y. Biondi et al.
eds. (1993). The Firm as an Entity: Implications for Economics, Accounting, and
Law. NY and London: Routledge (2007)

\bibitem {Tirole} J. Tirole, "On the Possibility of Speculation under Rational
Expectations," \textit{Econometrica}, 50, p. 1163-1181 (1982)

\bibitem {Fama2} E. F. Fama, "Efficient capital markets: a review of theory and
empirical work," \textit{Journal of Finance}, 25 (2), 383-417 (1970)

\bibitem {Fama3} E. F. Fama, \textit{Foundations of Finance}, New York: Basic Books (1976)

\bibitem {Fama1} E. F. Fama, The Behavior of Stock Market Prices, \textit{Journal
of Business}, 38 (1), p.\ 34-105 (1965)

\bibitem {Phelps} S. E. Phelps, \textquotedblleft Recent studies of speculative
markets in the controversy over rational expectations\textquotedblright,
European University Institute, WP n. 87/267, Florence (1987)

\bibitem {Frydman} R. Frydman, "Towards an understanding of market processes:
individual expectations, learning, and convergence to rational expectations
equilibrium," \textit{American Economic Review}, 72 (4), September, p. 652-668 (1982)

\bibitem {Hirota} S. Hirota and S. Sunder, \textquotedblleft Price bubbles sans dividend
anchors: Evidence from laboratory stock markets,\textquotedblright
\ \textit{Journal of Economic Dynamics and Control}, 31: 1875-1909  (2007)

\bibitem {Heemeijer} P. Heemeijer, C. Hommes, J. Sonnemans, and J. Tuinstra,
\textquotedblleft Price stability and volatility in markets with positive and
negative expectations feedback: An experimental
investigation,\textquotedblright\ \textit{Journal of Economic Dynamics and
Control}, 33, 1052--1072 (2009)

\bibitem {Biondi} Y. Biondi, and P. Giannoccolo, "Share Price Formation, Market
Exuberance and Accounting Design," Banque de France Foundation Research
Seminar, November 23, 2010. URL: http://ssrn.com/abstract=1690398 (2010)

\bibitem {Shiller} R. J. Shiller, \textit{Irrational Exuberance}.\ Princeton, N.J.:
Princeton University Press (2000)

\bibitem {LeRoy} S. F. LeRoy, "Rational Exuberance," \textit{Journal of Economic
Literature}, 42 (3) September: 783-804 (2004)



\end{thebibliography}
\end{document}